\documentclass[aps,amsmath,amssymb,pra,twocolumn,superscriptaddress,floatfix]{revtex4}
\usepackage{graphicx}
\usepackage{amsmath, amssymb, amsthm}
\usepackage{mathtools}
\usepackage{color}
\usepackage{braket}
\usepackage{hyperref}

\DeclareMathOperator{\polylog}{polylog}
\DeclareMathOperator\arctanh{arctanh}
\allowdisplaybreaks[4]

\theoremstyle{definition}

\theoremstyle{definition}

\theoremstyle{remark}

\begin{document}

\title{Cost-Reduced All-Gaussian Universality with the Gottesman-Kitaev-Preskill Code: Resource-Theoretic Approach to Cost Analysis}

\author{Hayata Yamasaki}
\email{yamasaki@qi.t.u-tokyo.ac.jp}
\affiliation{Photon Science Center, Graduate School of Engineering, The University of Tokyo, 7--3--1 Hongo, Bunkyo-ku, Tokyo 113--8656, Japan}

\author{Takaya Matsuura}
\email{matsuura@qi.t.u-tokyo.ac.jp}
\affiliation{Department of Applied Physics, Graduate School of Engineering, The University of Tokyo, 7--3--1  Hongo, Bunkyo-ku, Tokyo 113--8656, Japan}

\author{Masato Koashi}
\email{koashi@qi.t.u-tokyo.ac.jp}
\affiliation{Department of Applied Physics, Graduate School of Engineering, The University of Tokyo, 7--3--1  Hongo, Bunkyo-ku, Tokyo 113--8656, Japan}
\affiliation{Photon Science Center, Graduate School of Engineering, The University of Tokyo, 7--3--1 Hongo, Bunkyo-ku, Tokyo 113--8656, Japan}

\date{\today}

\begin{abstract}
  The Gottesman-Kitaev-Preskill (GKP) quantum error-correcting code has emerged as a key technique in achieving fault-tolerant quantum computation using photonic systems. Whereas~[Baragiola \textit{et al.}, Phys.\ Rev.\ Lett.\ \textbf{123}, 200502 (2019)] showed that experimentally tractable Gaussian operations combined with preparing a GKP codeword $\Ket{0}$ suffice to implement universal quantum computation, this implementation scheme involves a distillation of a logical magic state $\Ket{H}$ of the GKP code, which inevitably imposes a trade-off between implementation cost and fidelity. In contrast, we propose a scheme of preparing $\Ket{H}$ directly and combining Gaussian operations only with $\Ket{H}$ to achieve the universality without this magic state distillation. In addition, we develop an analytical method to obtain bounds of fundamental limit on transformation between $\Ket{H}$ and $\Ket{0}$, finding an application of quantum resource theories to cost analysis of quantum computation with the GKP code. Our results lead to an essential reduction of required non-Gaussian resources for photonic fault-tolerant quantum computation compared to the previous scheme.
\end{abstract}

\maketitle

\section{Introduction}
Photonic quantum systems provide promising architectures toward implementing quantum computation~\cite{C2,B8,doi:10.1063/1.5100160}.
Quantum computation brings advantages over conventional classical computation in terms of computational speedups~\cite{W5,H3,A3,B4} and stronger security~\cite{B14,B15}.
Compared to other matter-based candidates for implementing quantum computation such as superconducting qubits~\cite{W3,K8} and ion traps~\cite{H17,B3}, characteristics of the photonic architectures are scalability in generating quantum entanglement among more than one million optical modes~\cite{Y4} and flexibility in geometrical constraints on interactions that are essentially free from two-dimensional surface of the matter.
The scalability is especially key to attaining high fault tolerance in quantum computation by means of quantum error correction~\cite{G,D,T10,B}, where quantum information of a logical qubit is redundantly encoded in a physical quantum system.

To implement fault-tolerant quantum computation using photonic systems,  besides single-photon-based candidates such as Knill-Laflamme-Milburn scheme~\cite{knill2001scheme}, it is promising to exploit the Gottesman-Kitaev-Preskill (GKP) quantum error-correcting code~\cite{G1} for correcting errors that occur in continuous-variable (CV) systems~\cite{M1,F2,F6,N3,V4}.
The GKP code can encode a logical qubit into a CV degree of freedom in an optical mode.
Reference~\cite{B12} has recently shown that if we can realize a light source that emits an optical mode prepared in a codeword $\Ket{0}$ of a GKP code, experimentally tractable Gaussian operations combined with this light source suffice to implement universal quantum computation in a fault-tolerant way.
When we implement fault-tolerant quantum computation using qubits,
magic state distillation~\cite{B1} serves as a key technique for preparing a special type of logical state of a qubit-based quantum error-correcting code, \textit{a magic state} such as a Hadamard eigenstate $\Ket{\overline{H}}$ of the code, from noisy magic states.
Preparation of a codeword $\Ket{\overline{0}}$ of the qubit-based code is typically much easier, by means of projection using stabilizers~\cite{D}, than that of $\Ket{\overline{H}}$,
and cheap $\Ket{\overline{0}}$s can be combined with expensive $\Ket{\overline{H}}$s to achieve universal quantum computation in a fault-tolerant way.
The photonic scheme of fault-tolerant quantum computation in Ref.~\cite{B12} also exploits a magic state distillation for the GKP code, where many non-Gaussian $\Ket{0}$s of the GKP code are transformed by Gaussian operations into another GKP-code state $\Ket{H}$, \textit{a GKP magic state}.
This scheme suggests a route to implementing universal quantum computation by realizing \textit{only one type} of a GKP-code state $\Ket{0}$; that is, it is no longer required to develop technologies for realizing two different light sources for $\Ket{0}$ and $\Ket{H}$ of the GKP code and coordinating the two.
However, in contrast to the qubit-based codes, both $\Ket{0}$ and $\Ket{H}$ of the GKP code are non-Gaussian and hence costly to prepare compared to realizing Gaussian operations.
Thus, the overhead cost of consuming many expensive $\Ket{0}$s per distillation of $\Ket{H}$ may become a crucial obstacle in implementing quantum computation under this scheme.

To circumvent this obstacle arising from the magic state distillation and achieve a fundamental cost reduction in implementing photonic fault-tolerant quantum computation, this paper aims at putting forward an idea of preparing only the logical magic state $\Ket{H}$ of the GKP code.
We show a scheme that combines Gaussian operations only with $\Ket{H}$, instead of $\Ket{0}$, to implement universal quantum computation.
In contrast with the previous scheme, our scheme can be free from the overhead cost of the magic state distillation, because $\Ket{0}$ can be deterministically prepared from as few as two $\Ket{H}$s by means of the state-injection protocol~\cite{B1}.
While the state injection is well known in the qubit-based quantum computation, our key contribution is an essential reduction of non-Gaussian resources in implementing photonic fault-tolerant quantum computation.
Notice that the cost reduction stems from the nature of photonic architecture that non-Gaussian $\Ket{0}$ and $\Ket{H}$ of the GKP code are costly to prepare compared to performing Gaussian operations.
This cost reduction does not necessarily hold for qubit-based quantum error-correcting codes where logical $\Ket{0}$ is much easier to prepare than $\Ket{H}$;
it does not necessarily hold either for other architectures than the photonics, such as superconducting cavities~\cite{Campagne2019} and trapped-ion mechanical oscillators~\cite{Fluhmann2019} using its oscillator mode to prepare the GKP code, where Gaussian operations are not necessarily easy to implement compared to non-Gaussian operations.
In addition, we introduce a simple analytical method for obtaining a fundamental bound that limits transformation between $\Ket{H}$ and $\Ket{0}$ of the GKP code by any Gaussian operations, discovering an application of quantum resource theories~\cite{RevModPhys.91.025001} for CV quantum computation, especially the resource theory of non-Gaussianity~\cite{T1,A1}.
We also show feasibility of direct preparation of $\Ket{H}$.
The existing proposals~\cite{Pirandola2004,Vasconcelos2010,Etesse2014,Arrazola2018,Fabre2019,Travaglione2002,Motes2017,Weigand2018,Eaton2019,Tzitrin2019,Pirandola2006,Pirandola2006_2,Brooks2013,Terhal2016,Fluhmann2018,Lin2018,Campagne2019,Shi2019,Weigand2019,Fluhmann2019,Le2019} on realizing the GKP code mostly focus on the preparation of $\Ket{0}$, but we discuss generalizations of some of the proposals to demonstrate that $\Ket{H}$ can be prepared at a technological cost comparable to that of $\Ket{0}$ using these proposals.
Our results open up a previously overlooked yet arguably promising avenue toward implementing photonic fault-tolerant quantum computation by realizing a light source of $\Ket{H}$ of the GKP code rather than $\Ket{0}$.

The rest of this paper is structured as follows.
In Sec.~\ref{sec:gkp}, we recall conventions on the GKP code and universal quantum computation.
In Sec.~\ref{sec:state_injection}, we show a scheme of universal quantum computation that combines Gaussian operations only with $\Ket{H}$
In Sec.~\ref{sec:resource},
we present the resource-theoretical method for analyzing the fundamental limit on the transformations between $\Ket{H}$ and $\Ket{0}$ of the GKP code by Gaussian operations.
The feasibility of direct preparation of $\Ket{H}$ is shown in Sec.~\ref{sec:preparation}.
Our conclusion is given in Sec.~\ref{sec:conclusion}.

\section{\label{sec:gkp}Universal quantum computation using GKP qubits}

The GKP code~\cite{G1} is a CV code for encoding a logical qudit
into position quadrature $\hat{q}$ and momentum quadrature $\hat{p}$ of an oscillator, \textit{e.g.}, an optical mode at a physical level,
where we write $\hbar=1$, $\hat{q}\coloneqq\frac{1}{\sqrt{2}}\left(\hat{a}+\hat{a}^\dag\right)$, $\hat{p}\coloneqq\frac{1}{\sqrt{2}\mathrm{i}}\left(\hat{a}-\hat{a}^\dag\right)$, and $\hat{a}^\dag$ and $\hat{a}$ are creation and annihilation operators, respectively~\cite{C2,B8}.
Each of the logical codewords $\left\{\Ket{0},\Ket{1},\ldots\right\}$ of a GKP code is ideally a superposition of infinitely many eigenstates of $\hat{q}$.
The simplest class of the GKP codes is the one-mode square-lattice GKP code encoding one qubit per mode, and its logical codewords $\left\{\Ket{j}:j=0,1\right\}$ are represented as $\Ket{j}\propto\sum_{s\in\mathbb{Z}}\Ket{\sqrt{\pi}\left(2s+j\right)}_q$, where $\Ket{q_0}_q$ is an eigenstate of $\hat{q}$ satisfying $\hat{q}\Ket{q_0}_q=q_0\Ket{q_0}_q$.
In this paper, the GKP code refers to this square-lattice GKP code for simplicity of the presentation.
We refer to the logical qubit encoded in a physical mode by the GKP code as \textit{a GKP qubit}, and to a physical state of GKP qubits as \textit{a GKP state}.

While the codewords of the ideal GKP code are non-normalizable and hence unphysical, we can circumvent this normalization problem by considering an approximate GKP code, where a standard form of the approximate GKP codewords is given in Ref.~\cite{M10}.
While the eigenstates of $\hat{q}$ in the definition of the ideal GKP codewords can be considered to be infinitely squeezed,
the approximate GKP code has approximately orthogonal codewords $\left\{\Ket{j_{\sigma^2}}:j=0,1\right\}$ given by replacing each infinitely squeezed eigenstate of $\hat{q}$ in the definition of the ideal GKP codewords with a finitely squeezed vacuum state of variance $\sigma^2$ weighted by a Gaussian envelop~\cite{G1,M10}, as summarized in Appendix~\ref{sec:standard}.
By convention~\cite{M1,M10}, we represent the degree of the approximation using the squeezing level in decibel, \textit{i.e.}, $-10\log_{10}(2\sigma^2)$.
The approximate GKP codewords approach to the ideal ones as $\sigma\to 0$, that is, $-10\log_{10}(2\sigma^2)\to\infty$.

Universal quantum computation is achieved by implementing an arbitrary quantum circuit on qubits that is composed of Clifford gates and non-Clifford gates~\cite{N4}.
A Clifford gate refers to a quantum logic gate generated by the Hadamard gate $H$, the phase gate $S$, and the controlled NOT gate $CNOT$, while a non-Clifford gate otherwise, such as the $T$ gate.
Operations composed of preparing qubits in Pauli eigenstates, applying Clifford gates to the qubits, and measuring the qubits in Pauli eigenbases are called Clifford operations, which can implement only a subclass of quantum computation, and efficient classical simulation of Clifford operations is possible~\cite{A2}.
Clifford operations combined with non-Clifford gates, \textit{e.g.}, the $T$ gate, can achieve universal quantum computation~\cite{N4}.

We can implement most of the logical Clifford operations on GKP qubits by Gaussian operations~\cite{G1}.
Gaussian operations~\cite{C2,B8} are a subclass of operations on CV quantum systems composed of preparing the vacuum state, applying Gaussian unitary gates, and performing homodyne detection.
Gaussian operations are technologically easy to implement compared to non-Gaussian operations, but efficient classical simulation of Gaussian operations is possible.
If a pure CV state can be prepared only by Gaussian operations from the vacuum state, this pure state is a Gaussian state, that is, a CV state whose Wigner function is represented as a Gaussian function, where the Winger function of a density operator $\hat\psi$ is defined as $W_{\hat\psi}\left(q,p\right)\coloneqq\frac{1}{2\pi}\int_{-\infty}^{\infty}dx\,\mathrm{e}^{\mathrm{i}xp}\Braket{q-\frac{x}{2}|\hat{\psi}|q+\frac{x}{2}}$.
Logical Clifford gates on GKP qubits can be implemented by Gaussian operations achieving the following symplectic transformations of quadratures~\cite{G1}:
$H:\hat{q}\to\hat{p},\,\hat{p}\to-\hat{q}$;
$S:\hat{q}\to\hat{q},\,\hat{p}\to\hat{p}-\hat{q}$;
$CNOT:\hat{q}_1\to\hat{q}_1,\,\hat{p}_1\to\hat{p}_1-\hat{p}_2,\,\hat{q}_2\to\hat{q}_1+\hat{q}_2,\,\hat{p}_2\to\hat{p}_2$,
where $\hat{q}_1, \hat{p}_1$ and $\hat{q}_2, \hat{p}_2$ are quadratures of the control and target modes, respectively.
The measurement in the logical Pauli-$Z$ basis $\left\{\Ket{0},\Ket{1}\right\}$ of a GKP qubit can be implemented by homodyne detection for measuring the $\hat{q}$ quadrature of the mode.
However, we remark that Gaussian operations and logical Clifford operations on GKP qubits are different in that Pauli eigenstates of the GKP code, such as $\Ket{0}$ and $\Ket{1}$, are non-Gaussian; that is, initialization of GKP qubits requires non-Gaussian operations.

As for logical non-Clifford gates on GKP qubits, Ref.~\cite{G1} provides protocols for deterministically applying the $T$ gate to any GKP state using Gaussian operations and an auxiliary mode prepared either in a GKP Hadamard eigenstate $\Ket{H}\coloneqq\left(\cos\frac{\pi}{8}\right)\Ket{0}+\left(\sin\frac{\pi}{8}\right)\Ket{1}$, a GKP $\frac{\pi}{8}$ phase state $\Ket{\frac{\pi}{8}}\coloneqq\frac{1}{\sqrt{2}}\left(\mathrm{e}^{-\mathrm{i}\frac{\pi}{8}}\Ket{0}+\mathrm{e}^{\mathrm{i}\frac{\pi}{8}}\Ket{1}\right)$, or a cubic phase state.
A CV state that assists Gaussian operations to apply a logical non-Clifford gate to GKP qubits, such as $\Ket{H}$, is called \textit{a GKP magic state}.

\section{\label{sec:state_injection}Deterministic all-Gaussian universality using a GKP magic state}

Toward implementing fault-tolerant quantum computation using photonic systems, it is promising to combine Gaussian operations with GKP qubits~\cite{M1,F2,F6,N3,V4}.
This is because Gaussian operations by themselves cannot correct Gaussian errors that occur in CV photonic systems~\cite{N1}, but combining Gaussian operations with an approximate GKP code concatenated with a multiqubit quantum error-correcting code, we can achieve the quantum error correction for CV systems~\cite{M1}.

Which GKP state to prepare and how to combine the GKP state with Gaussian operations matter in reducing technological cost of implementing quantum computation, since GKP states are non-Gaussian.
Employing the non-Gaussianity of $\Ket{0}$ of a GKP qubit, Ref.~\cite{B12} has recently shown a protocol based on magic state distillation~\cite{B1,R1} that probabilistically and approximately transforms auxiliary GKP qubits prepared in $\Ket{0}\otimes\Ket{0}\otimes\cdots$ into a magic state $\Ket{H}$ only using Gaussian operations, whereas it is still unknown whether a deterministic or exact Gaussian transformation from a finite number of $\Ket{0}$s to $\Ket{H}$ is possible or not.
This protocol suggests that when Gaussian operations are available, a light source that can emit \textit{only a single type} of a GKP state $\Ket{0}$ suffices to implement universal quantum computation.
However, this protocol imposes the overhead implementation cost arising from the magic state distillation; that is, whenever we need to use one GKP magic state $\Ket{H}$ to implement one logical $T$ gate up to a sufficiently small accuracy $\epsilon$ in fidelity, the light source has to generate many $\Ket{0}$s.
This overhead cost per $T$ gate on a GKP qubit increases the total implementation cost of fault-tolerant quantum computation including that of implementing fault-tolerant logical non-Clifford gates on a qubit-based quantum error-correcting code that we concatenate with the GKP code.
In general, the overhead cost caused by the magic state distillation in terms of the number of auxiliary GKP qubits, which are prepared in $\Ket{0}$ in the case of the scheme in Ref.~\cite{B12}, amounts to~\cite{B2,J1,H2}
\begin{equation}
  \label{eq:distillation_cost}
  O\left(\polylog\left(\frac{1}{\epsilon}\right)\right)\quad\text{as }\epsilon\to 0.
\end{equation}

\begin{figure}[t]
  \centering
  \includegraphics[width=3.4in]{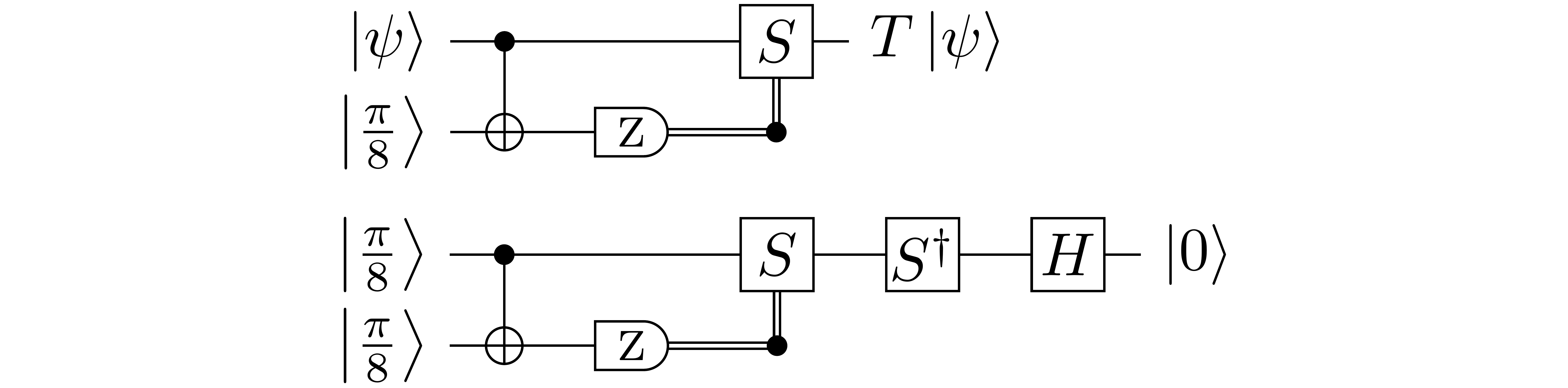}
  \caption{\label{fig:state_injection}A quantum circuit of state injection for applying the $T$ gate to any one-qubit input state $\Ket{\psi}$ by Clifford operations assisted by an auxiliary input qubit prepared in $\Ket{\frac{\pi}{8}}$ at the top, and that for converting a two-qubit input state $\Ket{\frac{\pi}{8}}^{\otimes 2}$ to $\Ket{0}$ at the bottom. The latter conversion circuit can be implemented only by adaptive Gaussian operations on GKP qubits, namely, Clifford gates ($CNOT$, $S$, $S^\dag$, and $H$) that are implemented with Gaussian unitary operations, and conditioning on a $Z$-basis measurement outcome that is implemented with a homodyne detection.}
\end{figure}

To reduce this cost of the required number of auxiliary GKP qubits, we here propose choosing $\Ket{H}$ instead of $\Ket{0}$ as the single GKP state for achieving universal quantum computation.
Since $\Ket{H}$ and $\Ket{\frac{\pi}{8}}$ are related by Clifford operations as $\Ket{H}=SH\Ket{\frac{\pi}{8}}$,
and we can implement any logical Clifford gates on the ideal GKP qubits by Gaussian operations, the following description of our proposal uses these states interchangeably.
Toward the cost reduction, recall a well-known quantum circuit for state injection~\cite{G1,B1} given at the top of Fig.~\ref{fig:state_injection}, which can apply the $T$ gate to an arbitrary one-qubit input state $\Ket{\psi}$ only using Clifford operations assisted by an auxiliary qubit prepared in $\Ket{\frac{\pi}{8}}$.
Inputting $\Ket{\psi}=\Ket{\frac{\pi}{8}}$ of a GKP qubit to this circuit and using additional Clifford gates,
we can deterministically transform two GKP qubits prepared in $\Ket{\psi}\otimes\Ket{\frac{\pi}{8}}=\Ket{\frac{\pi}{8}}^{\otimes 2}$ into $\Ket{0}$ only by Gaussian operations as shown at the bottom of Fig.~\ref{fig:state_injection}.
This protocol indicates that Gaussian operations combined with a light source of the GKP state $\Ket{H}$ can prepare $\Ket{0}$; that is, this combination can implement universal quantum computation.
This protocol is fault tolerant, \textit{i.e.}, can correct errors on CV systems as long as we use an approximate GKP code that approximates the ideal one sufficiently well, in the same way as the protocol in Ref.~\cite{B12}.
Our deterministic protocol using $\Ket{H}$s to prepare $\Ket{0}$ can be advantageous over the probabilistic protocol in Ref.~\cite{B12} using $\Ket{0}$s to prepare $\Ket{H}$; in contrast to~\eqref{eq:distillation_cost}, the overhead cost of the number of auxiliary GKP qubits, which are prepared in $\Ket{H}$ in our protocol, per preparation of $\Ket{0}$ is deterministically bounded by a practically small constant, \textit{i.e.},
\begin{equation}
  \label{eq:constant_cost}
  2=O\left(1\right),
\end{equation}
where $\Ket{0}$ is exactly ($\epsilon=0$) obtained in the ideal case.

\section{\label{sec:resource}A resource-theoretical framework for analyzing fundamental limitations in GKP state conversion}

Since transformation between GKP states $\Ket{H}$ and $\Ket{0}$ under Gaussian operations is crucial in implementing quantum computation by Gaussian operations with only one of $\Ket{H}$ and $\Ket{0}$,
we here develop a simple analytical method for obtaining fundamental bounds of the convertibility of the GKP states.
Our analysis is based on the resource theory of non-Gaussianity, where Gaussian operations are considered to be free and non-Gaussianity is regarded as a resource for assisting Gaussian operations~\cite{T1,A1,L2,Z1}.
Following Refs.~\cite{T1,A1}, we include adaptive Gaussian operations conditioned on measurement outcomes of homodyne detection in the free operations.
Note that while Gaussian operations on GKP qubits are analogous to Clifford operations of qubits, the resource theories of magic~\cite{V1,H1} using the Clifford operations as the free operations are insufficient for our analysis.
This is because Gaussian operations cannot prepare Pauli eigenstates of GKP qubits, \textit{e.g.}, $\Ket{0}$, but in the resource theory of magic, Pauli eigenstates of qubits are free states and can be prepared arbitrarily.

To analyze the convertibility between GKP states under Gaussian operations, we can use a measure that quantifies non-Gaussianity of a CV state.
One way to quantify the non-Gaussianity of a given CV state $\hat{\psi}$ is to use the negativity~\cite{T1,A1} of the Wigner function $W_{\hat{\psi}}$ of $\hat{\psi}$ defined as
$N\left(\hat{\psi}\right)\coloneqq\int_{-\infty}^{\infty}dq\int_{-\infty}^{\infty}dp\,\left|W_{\hat{\psi}}\left(q,p\right)\right|$,
where the normalization of $\hat\psi$ yields $\int_{-\infty}^{\infty}dq\int_{-\infty}^{\infty}dp\,W_{\hat{\psi}}\left(q,p\right)=1$.
Note that $\ln N$ yields the logarithmic negativity used in Refs.~\cite{T1,A1}.
The negativity $N$ does not increase under any Gaussian operations (\textit{i.e.}, $N$ has monotonicity).

\begin{figure}[t]
    \centering
    \includegraphics[width=3.4in]{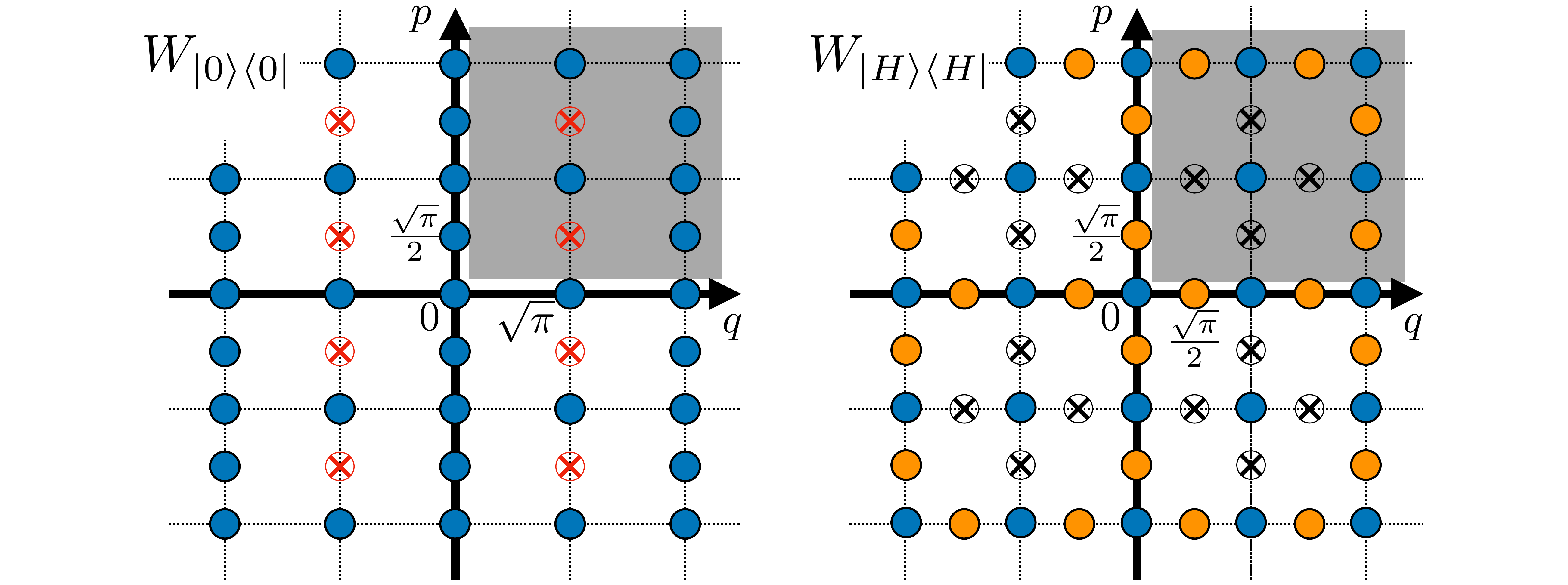}
    \caption{\label{fig:wigner}Wigner functions of ideal GKP states $\Ket{0}$ on the left and $\Ket{H}$ on the right, where each blue filled circle represents a positive delta function $\delta$, each red circled X represents a negative delta function $-\delta$, each yellow filled circle represents a weighted positive delta function $\frac{1}{\sqrt{2}}\delta$, and each black circled X represents a weighted negative delta function $-\frac{1}{\sqrt{2}}\delta$, \textit{e.g.}, $W_{\Ket{0}\Bra{0}}(q,p)\propto\sum_{s,t\in\mathbb{Z}}{(-1)}^{st}\delta\left(q-\sqrt{\pi}s\right)\delta\left(p-\frac{\sqrt{\pi}}{2}t\right)$ up to normalization. These Wigner functions have periodicity, where the gray region shows a period.}
\end{figure}

We here put forward a simple analytical method for calculating the negativities of ideal GKP states to compare their non-Gaussianity.
The Wigner functions of ideal GKP states consist of infinitely many Dirac delta functions that are arranged according to a square lattice, as depicted in Fig.~\ref{fig:wigner}.
Since the ideal GKP states are non-normalizable,
the negativity $N$ of an ideal GKP state is not well-defined.
To circumvent this mathematical subtlety, we exploit the periodicity of the Wigner functions shown in Fig.~\ref{fig:wigner} and evaluate the negativity of an ideal GKP state by replacing the improper integral of $N$ from $-\infty$ to $\infty$ with an integral over one period.
In particular, in place of $N$, we define
\begin{equation}
  \label{eq:neg_unit_cell}
  \widetilde{N}\left(\hat\psi\right)\coloneqq\frac{\int_\mathcal{I} dq\,\int_\mathcal{I} dp\,\left|W_{\hat{\psi}}\left(q,p\right)\right|}{\int_\mathcal{I} dq\,\int_\mathcal{I} dp\,W_{\hat{\psi}}\left(q,p\right)},
\end{equation}
where $\mathcal{I}\coloneqq\left[0+\epsilon,2\sqrt{\pi}+\epsilon\right]$ for any fixed $\epsilon\in\left(0,\frac{\sqrt{\pi}}{2}\right)$ represents the period shown in Fig.~\ref{fig:wigner}, and the denominator is chosen so that we have $\widetilde{N}\left(\hat{\psi}\right)=1$ for a state $\hat\psi$ that has a nonnegative Wigner function.
Then, by counting delta functions in Fig.~\ref{fig:wigner}, we obtain
\begin{align}
  \label{eq:zero}
  \widetilde{N}(\Ket{0}\Bra{0})&=\frac{8}{4}=2,\\
  \label{eq:h}
  \widetilde{N}(\Ket{H}\Bra{H})&=\frac{4 + 8\times\left(1/\sqrt{2}\right)}{4} = 1+\sqrt{2}.
\end{align}
Thus, we quantitatively compare the non-Gaussianity of $\Ket{H}$ and $\Ket{0}$ by
\begin{equation}
  \label{eq:bound}
  \frac{\widetilde{N}\left(\Ket{H}\Bra{H}\right)}{\widetilde{N}\left(\Ket{0}\Bra{0}\right)}=\frac{1+\sqrt{2}}{2}>1,
\end{equation}
which implies that $\Ket{H}$ has more non-Gaussianity than $\Ket{0}$, and hence no Gaussian operation can deterministically transform $\Ket{0}$ into $\Ket{H}$.

\begin{figure}[t]
  \centering
  \includegraphics[width=3.4in]{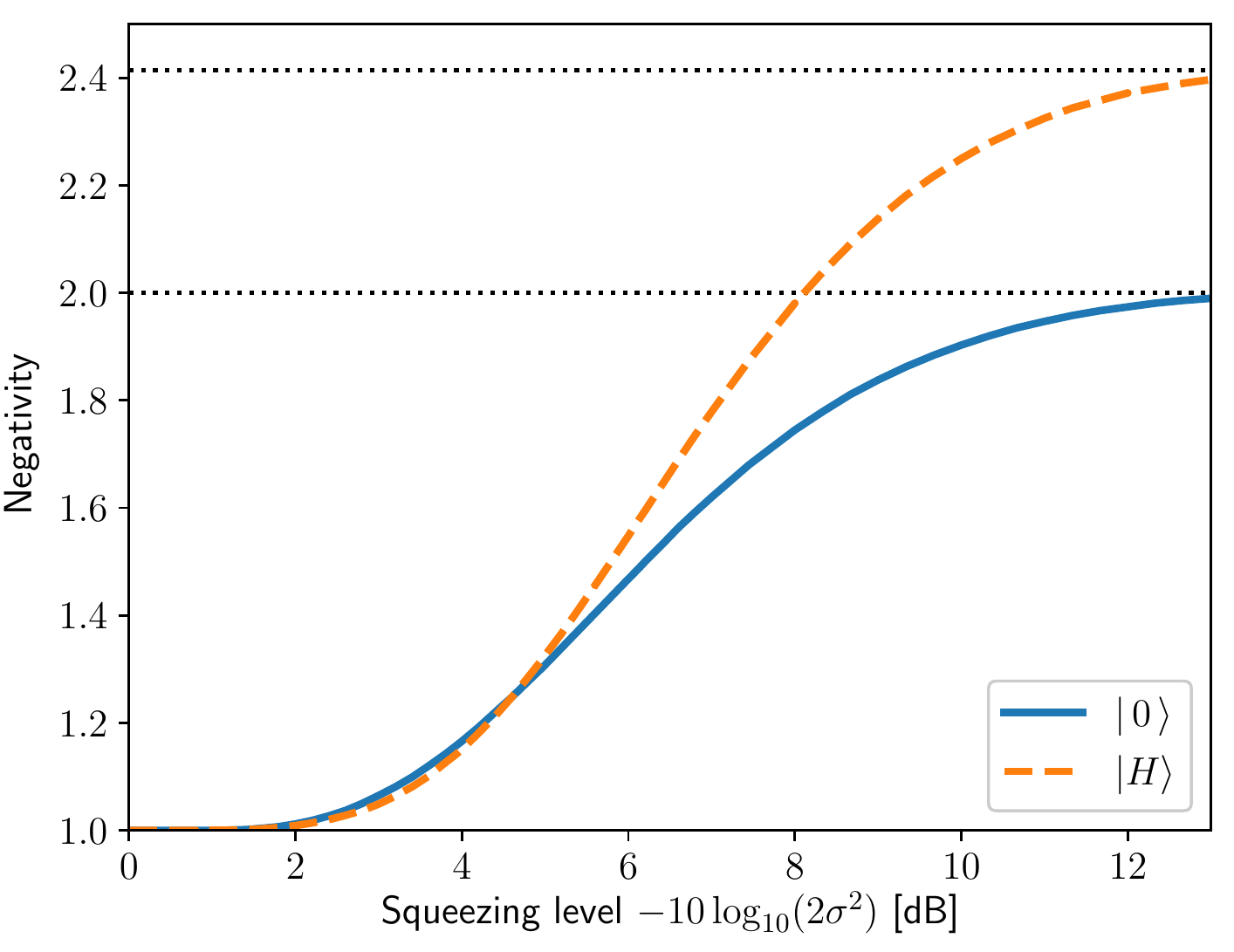}
  \caption{\label{fig:negativity}Negativities of the Wigner functions of $\Ket{0_{\sigma^2}}$ (blue solid line) and $\Ket{H_{\sigma^2}}\propto\left(\cos\left(\frac{\pi}{8}\right)\Ket{0_{\sigma^2}} + \sin\left(\frac{\pi}{8}\right) \Ket{1_{\sigma^2}} \right)$ (orange dashed line) with respect to the squeezing level $-10\log_{10}(2\sigma^2)$.  Negativity of $\Ket{0_{\sigma^2}}$ approaches to $2$, and that of $\Ket{H_{\sigma^2}}$ to $1+\sqrt{2}=2.41\cdots$, as expected from the calculations~\eqref{eq:zero} and~\eqref{eq:h}.}
\end{figure}

To justify using $\widetilde{N}$ as a substitute of $N$ for ideal GKP states, we also perform a numerical calculation of the negativity $N$ of approximate GKP states, which is well-defined.
Figure~\ref{fig:negativity} shows the negativities of the Wigner functions of $\Ket{0_{\sigma^2}}$ and $\Ket{H_{\sigma^2}}\propto\left(\cos\left(\frac{\pi}{8}\right)\Ket{0_{\sigma^2}} + \sin\left(\frac{\pi}{8}\right) \Ket{1_{\sigma^2}} \right)$ with respect to the squeezing level $-10\log_{10}(2\sigma^2)$ of the approximate codewords $\Ket{j_{\sigma^2}}$.
For the plot, we performed the numerical integration for the absolute values of the Wigner functions of $\Ket{0_{\sigma^2}}$ and $\Ket{H_{\sigma^2}}$ using Mathematica 11.2.0.
The figure indicates that the negativity of $\Ket{0_{\sigma^2}}$ approaches to $\widetilde{N}\left(\Ket{0}\Bra{0}\right)=2$ as $-10\log_{10}(2\sigma^2)\to\infty$, that is, in the limit of good approximation, and that of $\Ket{H_{\sigma^2}}$ to $\widetilde{N}\left(\Ket{H}\Bra{H}\right)=1+\sqrt{2}$, as expected from our arguments.

While a Gaussian transformation from $\Ket{H}^{\otimes 2}$ to $\Ket{0}$ is achievable as shown in Fig.~\ref{fig:state_injection},
our method for evaluating the negativities can conversely provide an upper bound in generating $\Ket{0}$s of GKP qubits from $\Ket{H}$ by Gaussian operations.
In the same way as the multiplicativity of the negativity $N$~\cite{T1,A1}, $\widetilde{N}$ is multiplicative, \textit{i.e.}, $\widetilde{N}(\hat{\psi}^{\otimes n})={(\widetilde{N}(\hat{\psi}))}^n$.
The multiplicativity of $\widetilde{N}$ shows that $\Ket{H}^{\otimes 2}$ cannot be transformed into $\Ket{0}^{\otimes 3}$ by any Gaussian operation because we have
\begin{equation}
  \frac{\widetilde{N}(\Ket{H}\Bra{H}^{\otimes 2})}{\widetilde{N}(\Ket{0}\Bra{0}^{\otimes 3})}=\frac{{(1+\sqrt{2})}^2}{2^3}<1.
\end{equation}

Our method finds a useful application of quantum resource theory for obtaining fundamental bounds that limit GKP state conversion.
Note that the evaluation of the negativity of the ideal (infinitely squeezed) GKP states was also discussed in Ref.~\cite{G2}, which is done independently of our work.
Yet in contrast to Ref.~\cite{G2}, our numerical calculation shows the negativity of GKP states not only at the infinite squeezing level but all the squeezing levels in Fig.~\ref{fig:negativity}, justifying the calculation of the negativity in the limit of the infinite squeezing level.
Furthermore, our crucial contribution is to apply this calculation of the negativity to obtaining the fundamental limit on transformation between GKP states, from the resource-theoretical perspective.
While the monotonicity of the negativity by itself may provide few implications for the achievability, our analysis raises the following open questions for future research.
First, it remains unclear whether there exist Gaussian operations that transform one copy of $\Ket{H}$ to $\Ket{0}$, and how we can achieve such a Gaussian transformation if it exists.
Second, whereas the probabilistic transformation from multiple $\Ket{0}$s to $\Ket{H}$ is used in the scheme of Ref.~\cite{B12}, whether a deterministic transformation of a finite number of $\Ket{0}$s to $\Ket{H}$ is possible or not is still open; in particular, the calculations~\eqref{eq:zero} and~\eqref{eq:h} of the negativity do not prohibit deterministic transformation from $\Ket{0}^{\otimes 2}$ to $\Ket{H}$, and further research is needed to conclude the feasibility of this transformation.
Lastly, it would be interesting to investigate whether Gaussian operations combined with post-selection can increase the negativity of a GKP state with nonzero probability; \textit{e.g.}, it is interesting to investigate whether there can be a Gaussian protocol for transforming only one copy of $\Ket{0}$ into $\Ket{H}$ with nonzero probability in the limit of good approximation of the GKP code.
Our resource-theoretical arguments and methods open a starting point for tackling these types of questions on the GKP state conversion.

\section{\label{sec:preparation}Feasibility of preparing a GKP magic state}

Since Gaussian operations combined with a GKP magic state $\Ket{H}$ can be advantageous in implementing quantum computation over those with $\Ket{0}$ of GKP qubits, we here discuss possible protocols for preparing $\Ket{H}$.
Our following discussion is based on the proposals for the photonic implementation of approximate GKP qubits~\cite{Travaglione2002,Pirandola2004,Vasconcelos2010,Etesse2014,Motes2017,Arrazola2018,Weigand2018,Eaton2019,Fabre2019,Tzitrin2019},
while there also exist other proposals and experimental demonstrations of generating approximate GKP codewords in various systems~\cite{Pirandola2006,Pirandola2006_2,Brooks2013,Terhal2016,Fluhmann2018,Lin2018,Campagne2019,Shi2019,Weigand2019,Fluhmann2019,Le2019}.
Note that architectures such as superconducting cavities~\cite{Campagne2019} and trapped-ion mechanical oscillators~\cite{Fluhmann2019} are also promising candidates to realize the GKP code, but we here focus on photonic implementations since Gaussian operations are not necessarily easier to implement than non-Gaussian operations on the superconducting cavities and the trapped-ion mechanical oscillators.
We remark that these existing proposals mostly focus on preparing $\Ket{0}$ or $\Ket{1}$ of the GKP code.
Some of the proposals, such as those in Refs.~\cite{Pirandola2004,Vasconcelos2010,Etesse2014,Weigand2018,Arrazola2018,Fabre2019}, may not be suitable for the direct preparation of $\Ket{H}$ as we discuss in Appendix~\ref{sec:not}
In contrast, the protocols for the GKP state preparation proposed in Refs.~\cite{Travaglione2002,Motes2017,Weigand2018,Eaton2019,Tzitrin2019} can be easily modified for preparing $\Ket{H}$.

Two promising routes toward preparing $\Ket{H}$ directly in photonic systems are to use interaction between a discrete-variable system and an optical mode in the cavity QED setups~\cite{Travaglione2002,Motes2017,Weigand2018}, and to use linear optical circuits followed by photon-number-resolving (PNR) detectors~\cite{Eaton2019,Tzitrin2019}.
References~\cite{Travaglione2002,Motes2017,Weigand2018} consider an interaction between qudits and an optical mode in the cavity QED setups; Refs.~\cite{Travaglione2002,Weigand2018} use a recursive application of controlled-displacement operator between a qubit and an optical mode, while Ref.~\cite{Motes2017} utilizes a spin-$J$ system, \textit{i.e.}, a qudit instead of the qubit, prepared in a spin coherent state.
Although these protocols were aimed at preparing $\Ket{0}$ or $\Ket{1}$, the protocols can be modified for preparation of $\Ket{H}$ if we can perform an additional non-Clifford measurement on the qubit or qudit as we discuss in Appendix~\ref{sec:qubit}.
References~\cite{Eaton2019,Tzitrin2019} consider generating non-Gaussian states using linear optical circuits followed by PNR detectors.
In contrast with other protocols, this protocol affords implementations of $\Ket{0}$ and $\Ket{H}$ on an equal footing with almost the same resource requirements, as pointed out in Ref.~\cite{Tzitrin2019}.
Furthermore, preparing only one type of GKP state in this protocol may be desired since optical circuits and PNR detectors to generate the GKP state need to be finely tuned to keep the fidelity high.
These protocols indicate that it is feasible to prepare $\Ket{H}$ of GKP qubits with a technological requirement comparable to preparing $\Ket{0}$.

\section{\label{sec:conclusion}Conclusion}

We have proposed a photonic scheme of implementing universal quantum computation in a fault-tolerant way, where Gaussian operations are combined with a light source emitting the GKP magic state $\Ket{H}$, rather than the GKP codeword $\Ket{0}$ in the previous scheme of Ref.~\cite{B12}.
Our main contribution is the essential reduction of non-Gaussian resources, \textit{i.e.}, the number of GKP qubits, in implementing the computation, achieved by the direct preparation of $\Ket{H}$ in place of $\Ket{0}$ of the GKP code for avoiding the magic state distillation.
This cost reduction in the photonic quantum computation using the GKP code is a result of its intrinsic property that both $\Ket{H}$ and $\Ket{0}$ of the GKP code are costly to prepare, which does not necessarily hold for an error-correcting code for qubits.
In contrast with the previous scheme using $\Ket{0}$, our scheme can be free from the overhead cost given by~\eqref{eq:distillation_cost} of the magic state distillation for preparing $\Ket{H}$ from $\Ket{0}$s, and achieves as small as a constant overhead cost given by~\eqref{eq:constant_cost} in preparing $\Ket{0}$ from $\Ket{H}$s.

Our results put forward an argument on which of the two possible light sources of GKP states, $\Ket{H}$ or $\Ket{0}$, to realize toward implementing photonic fault-tolerant quantum computation, while a more concrete cost estimation of these two may require further assumptions on advances in photonic technologies and hence is left for future work.
In addition to constructing the cost-reduced scheme,
we have also introduced an analytical technique for addressing fundamental limitations in transformation between GKP states $\Ket{H}$ and $\Ket{0}$ under any Gaussian operations.
This technique, based on the resource theory of non-Gaussianity, discovers an application of quantum resource theories to quantum computation implemented by CV systems, progressing beyond applications of the resource theory of magic to that implemented by discrete-variable systems.
We have also discussed two possible protocols for directly preparing a photonic system in $\Ket{H}$; one is based on Refs.~\cite{Travaglione2002,Motes2017,Weigand2018}, and the other on Refs.~\cite{Eaton2019,Tzitrin2019}.
We point out here that not much attention has been paid to direct preparation of the GKP magic state $\Ket{H}$.
Our proposal and results open up future research on these lines to explore more efficient use of the GKP magic state $\Ket{H}$ and its preparation methods toward the goal of realizing photonic fault-tolerant quantum computation.

\acknowledgments{This work was supported by CREST (Japan Science and Technology Agency) JPMJCR1671 and Cross-ministerial Strategic Innovation Promotion Program (SIP) (Council for Science, Technologyand Innovation (CSTI)).}

\bibliography{citation_bibtex}

\appendix%

\section{\label{sec:standard}Standard form of approximate GKP codewords}

We review a standard form of approximate GKP codewords proposed in Ref.~\cite{M10}.
For approximate codewords $\{\Ket{j_{\sigma^2}}:j=0,1\}$, we use the following standard form~\cite{M10}:
\begin{widetext}
\begin{align}
  \Ket{j_{\sigma^2}} &= \frac{1}{{(1/4-\sigma^4)}^{\frac{1}{4}}\sqrt{N_{\sigma^2,j}}}\; \mathrm{e}^{-\left(\arctanh(2\sigma^2)\right)\left(\hat{a}^{\dagger}\hat{a} + \frac{1}{2}\right)}\Ket{j^{\left(\mathrm{ideal}\right)}}\\
                     &=\frac{1}{\sqrt{\sqrt{\pi}\sigma^2 N_{\sigma^2,j}}}\sum_{s=-\infty}^{\infty}\int_{-\infty}^{\infty}{dq}\,\mathrm{e}^{-\frac{2\sigma^2}{2\left(1-4\sigma^4\right)}{\left(\left(2s+j\right)\sqrt{\left(1-4\sigma^4\right)\pi}\right)}^2}\mathrm{e}^{-\frac{1}{2\left(2\sigma^2\right)}{\left(q-\left(2s+j\right)\sqrt{\left(1-4\sigma^4\right)\pi}\right)}^2}\Ket{q}_q,
\end{align}
\end{widetext}
where $\Ket{j^{\left(\mathrm{ideal}\right)}}\coloneqq\sqrt{2\sqrt{\pi}}\sum_{s\in\mathbb{Z}}\Ket{\sqrt{\pi}\left(2s+j\right)}_q$ represents the ideal GKP codewords, $N_{\sigma^2,j}$ is a constant for normalization, and the factor $\mathrm{e}^{-\frac{2\sigma^2}{2\left(1-4\sigma^4\right)}{\left(\left(2s+j\right)\sqrt{\left(1-4\sigma^4\right)\pi}\right)}^2}$ and the state $\int_{-\infty}^{\infty}{dq}\,\mathrm{e}^{-\frac{1}{2\left(2\sigma^2\right)}{\left(q-\left(2s+j\right)\sqrt{\left(1-4\sigma^4\right)\pi}\right)}^2}\Ket{q}_q$ can be regarded as the Gaussian envelop and the finitely squeezed vacuum state, respectively.
The Hadamard eigenstate $\Ket{H_{\sigma^2}}$ of the approximate GKP code is given by
\begin{align}
  &\Ket{H_{\sigma^2}}=\nonumber\\
  &\frac{1}{\sqrt{1 + \frac{1}{\sqrt{2}}\Re\left(\Braket{0_{\sigma^2}|1_{\sigma^2}}\right)}} \left(\cos\left(\frac{\pi}{8}\right)\Ket{0_{\sigma^2}} + \sin\left(\frac{\pi}{8}\right) \Ket{1_{\sigma^2}} \right),
\end{align}
where $\Re$ represents the real part, and the prefactor comes from the fact that $\Ket{0_{\sigma^2}}$ and $\Ket{1_{\sigma^2}}$ have non-zero overlap $\Braket{0_{\sigma^2}|1_{\sigma^2}}\neq 0$.

The normalization factor $N_{\sigma^2,j}$, the overlap $\Braket{0_{\sigma^2}|1_{\sigma^2}}$, and the Wigner functions of $\Ket{j_{\sigma^2}}$ and $\Ket{H_{\sigma^2}}$ can be obtained from the results in Ref.~\cite{M10}.
To show them here, we define the theta function with rational characteristics $(a,b)$ as
\begin{equation}
  \vartheta \! \left[ \begin{subarray}{c} a \\ \ \\ b \end{subarray}\right] \! \left(z,\tau\right)\coloneqq  \sum_{s\in\mathbb{Z}}\exp[\pi \mathrm{i} \tau {(s+a)}^2 + 2\pi \mathrm{i} (z+b) (s+a)].
\end{equation}
Then, Ref.~\cite{M10} shows that the normalization factor $N_{\sigma^2,j}$ and the overlap $\Braket{0_{\sigma^2}|1_{\sigma^2}}$ are given respectively by
\begin{widetext}
\begin{align}
  N_{\sigma^2,j} &= \vartheta \! \left[ \begin{subarray}{c} \frac{j}{d} \\ \ \\ 0 \end{subarray}\right] \! \left(0,8\mathrm{i}\sigma^2 \right) \vartheta \! \left[ \begin{subarray}{c} 0 \\ \ \\ 0 \end{subarray}\right] \! \left(0,\frac{\mathrm{i} \sigma^2 }{2}\right)
 + \vartheta \! \left[ \begin{subarray}{c} \frac{j}{d} + \frac{1}{2} \\ \ \\ 0 \end{subarray}\right] \! \left(0,8\mathrm{i}\sigma^2 \right) \vartheta \! \left[ \begin{subarray}{c} 0 \\ \ \\ \frac{1}{2} \end{subarray}\right] \! \left(0,\frac{\mathrm{i} \sigma^2}{2}\right),\\
 \Braket{0_{\sigma^2}|1_{\sigma^2}} &= \frac{1}{\sqrt{N_{\sigma^2,0}N_{\sigma^2,1}}} \left[ \vartheta \! \left[ \begin{subarray}{c} \frac{1}{4} \\ \ \\ 0 \end{subarray}\right] \! \left(0,8\mathrm{i}\sigma^2 \right) \vartheta \! \left[ \begin{subarray}{c} 0 \\ \ \\ \frac{1}{4} \end{subarray}\right] \! \left(0,\frac{\mathrm{i} \sigma^2 }{2}\right) + \vartheta \! \left[ \begin{subarray}{c} \frac{3}{4} \\ \ \\ 0 \end{subarray}\right] \! \left(0,8\mathrm{i}\sigma^2 \right) \vartheta \! \left[ \begin{subarray}{c} 0 \\ \ \\ \frac{3}{4} \end{subarray}\right] \! \left(0,\frac{\mathrm{i} \sigma^2 }{2}\right) \right].
\end{align}
\end{widetext}
Regarding the Wigner functions of $\Ket{0_{\sigma^2}}$ and $\Ket{H_{\sigma^2}}$, in this paper, we only need to show the Wigner representations of the operator $\Ket{j_{\sigma^2}}\Bra{j'_{\sigma^2}}$ for $j,j'\in\{0,1\}$.
Applying Lemma 1 in Ref.~\cite{M10} to the Wigner representation of $\Ket{j_{\sigma^2}}\Bra{j'_{\sigma^2}}$ in Proposition 2 in Ref.~\cite{M10}, we obtain
\begin{widetext}
\begin{align}
&W_{\Ket{j_{\sigma^2}}\Bra{j'_{\sigma^2}}}(q,p) \nonumber \\
&= \frac{1}{2\sigma^2\sqrt{N_{\sigma^2,j}N_{\sigma^2,j'}}} \left[ G_{\frac{1}{4\sigma^2}}(q)\; \vartheta \! \left[ \begin{subarray}{c} 0 \\ \ \\ \frac{j + j'}{4} \end{subarray}\right] \! \left(-\frac{q\sqrt{1-4\sigma^4}}{2\sqrt{\pi}},\frac{\mathrm{i}\sigma^2}{2} \right)\, G_{\frac{1}{4\sigma^2}}(p)\; \vartheta \! \left[ \begin{subarray}{c} \frac{j - j'}{4} \\ \ \\ 0 \end{subarray}\right] \! \left(-\frac{2p\sqrt{1-4\sigma^4}}{\sqrt{\pi}},8\mathrm{i}\sigma^2 \right) \right. \nonumber \\
&\hspace{3.5cm} \left. +\, G_{\frac{1}{4\sigma^2}}(q)\; \vartheta \! \left[ \begin{subarray}{c} 0 \\ \ \\ \frac{j + j'}{4} + \frac{1}{2} \end{subarray}\right] \! \left(-\frac{q\sqrt{1-4\sigma^4}}{2\sqrt{\pi}},\frac{\mathrm{i}\sigma^2}{2} \right)\, G_{\frac{1}{4\sigma^2}}(p)\; \vartheta \! \left[ \begin{subarray}{c} \frac{j - j'}{4} + \frac{1}{2} \\ \ \\ 0 \end{subarray}\right] \! \left(-\frac{2p\sqrt{1-4\sigma^4}}{\sqrt{\pi}},8\mathrm{i}\sigma^2 \right) \right],
\end{align}
\end{widetext}
where $G_{\frac{1}{4\sigma^2}}(x)$ denotes a probability density function of the normal distribution with variance $\frac{1}{4\sigma^2}$ given by
\begin{equation}
G_{\frac{1}{4\sigma^2}}(x)\coloneqq \sqrt{\frac{2\sigma^2}{\pi}} \mathrm{e}^{-2\sigma^2x^2}.
\end{equation}
Using these formulas for explicitly representing the approximate GKP states in terms of the theta function, we can evaluate the negativities $N\left(\Ket{0_{\sigma^2}}\Bra{0_{\sigma^2}}\right)$ and $N\left(\Ket{H_{\sigma^2}}\Bra{H_{\sigma^2}}\right)$ by numerically computing the integral of the theta function, as we show in the main text.

\section{\label{sec:not}Protocols that cannot straightforwardly prepare a GKP magic state}

We summarize existing protocols that can prepare a GKP codeword $\Ket{0}$ or $\Ket{1}$ but do not straightforwardly generalize to those for preparing a GKP magic state $\Ket{H}$.
Toward generating GKP codewords, Ref.~\cite{Pirandola2004} considers using the cross-Kerr non-linearity to couple two optical modes initially prepared in a coherent state and a squeezed coherent state respectively, followed by performing homodyne measurement of the mode initialized as the coherent state, which results in generating approximate GKP codewords $\Ket{0}$ or $\Ket{1}$.
In this scheme, $\Ket{H}$ cannot be directly prepared as long as Gaussian states are fed into the cross-Kerr interaction followed by the homodyne detection.
References~\cite{Vasconcelos2010,Etesse2014,Weigand2018} consider protocols that breed approximate GKP codewords from squeezed cat states.  A rough sketch of the protocol is that two premature GKP codewords, which are initially the even squeezed cat states, are interfered by a 50:50 beam-splitter, and then one of the modes is measured by a homodyne detector.  With a post-selection or a feed-back operation, the state becomes a better GKP codeword, that is, a superposition of the squeezed coherent states (approximately) weighted by a Gaussian function.  This scheme naturally prepares $\Ket{0}$ but does not prepare $\Ket{H}$, because a coherent superposition of $\Ket{0}$ and $\Ket{1}$ cannot be implemented with a naive application of the protocol.
Note that in addition to the breeding protocol, Ref.~\cite{Weigand2018} shows a protocol for preparing GKP codewords based on interaction between a qubit and an optical mode, and this protocol can be used for preparing a GKP magic state as we will show in Appendix~\ref{sec:qubit}.
As for other proposals, Ref.~\cite{Arrazola2018} analyzes optimization of parametrized non-Gaussian optical circuits by machine learning, and Ref.~\cite{Fabre2019} uses time-frequency degrees of freedom.
These proposals do not fit our current settings for preparing photonic GKP qubits where Gaussian operations are easy compared to non-Gaussian operations, while they are also interesting research directions.

\section{\label{sec:qubit}The preparation of a GKP magic state with a non-Clifford measurement on a qubit}

\begin{figure}[t]
  \centering
  \includegraphics[width=3.4in]{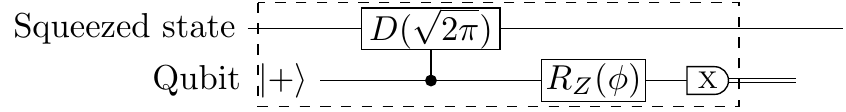}
  \caption{\label{fig:phase_estimation}A phase-estimation-like protocol for generating approximate GKP codewords proposed in Ref.~\cite{Weigand2018}, which is a refinement of the protocol in Ref.~\cite{Travaglione2002}.  The operations surrounded by the dashed line is recursively performed, while the parameter $\phi$ of the rotation around Z-axis $R_Z(\phi)$ is chosen according to the former outcomes of the measurement on the qubits.}
\end{figure}

\begin{figure}[t]
    \centering
    \includegraphics[width=3.4in]{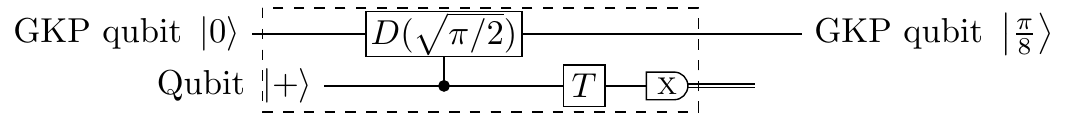}
    \caption{\label{fig:pi_over_8}A quantum circuit for preparing a GKP $\frac{\pi}{8}$ phase state $\Ket{\frac{\pi}{8}}$ from an input GKP codeword $\Ket{0}$ using the same setup as that in Ref.~\cite{Weigand2018}.  Note that $T$ gate is a rotation around $Z$-axis in the same way as $R_Z(\phi)$ in Fig.~\ref{fig:phase_estimation}, and thus this quantum circuit can be implemented using the same experimental setup as that in Fig.~\ref{fig:phase_estimation}.}
\end{figure}

Here we discuss how to prepare the GKP $\frac{\pi}{8}$ phase state $\Ket{\frac{\pi}{8}}$ using the interaction between a discrete-variable system and an optical mode based on the protocols in Refs.~\cite{Travaglione2002,Motes2017,Weigand2018}.
These protocols use a controlled-displacement gate, where we write a displacement operator as
\begin{align}
  D\left(\alpha\right)&\coloneqq \mathrm{e}^{\alpha {\hat{a}}^\dag - \alpha^\ast \hat{a}},\quad\alpha\in\mathbb{C},\\
  D\left(r\right)\Ket{q_0}_q&=\Ket{q_0+\sqrt{2}r}_q,\quad r\in\mathbb{R}.
\end{align}
Note that if we are allowed to use an interaction between a qubit and a photonic system beyond the controlled-displacement gate, an additional controlled-Fourier operation between the qubit and the photonic system can also prepare $\Ket{H}$ from $\Ket{0}$ as shown in Ref.~\cite{G1}, while the protocols in Refs.~\cite{Travaglione2002,Motes2017,Weigand2018} do not require this additional interaction.
For simplicity, we focus on the protocol proposed in Refs.~\cite{Travaglione2002,Weigand2018}, which recursively performs the controlled-displacement operation between a qubit and an optical mode, while a similar strategy to that in the following discussion is also applicable to the protocol proposed in Ref.~\cite{Motes2017}.
Figure~\ref{fig:phase_estimation} shows a protocol given in Ref.~\cite{Weigand2018}, which is a modification of the protocol in Ref.~\cite{Travaglione2002} while these protocols work essentially in the same way.
Using this protocol, we can prepare a superposition of a squeezed coherent state weighted by a Gaussian envelope.
Then, using the same experimental setup with a modification of parameters, we can obtain the GKP $\frac{\pi}{8}$ phase state $\Ket{\frac{\pi}{8}}$ from an input GKP codeword $\Ket{0}$, as shown in Fig.~\ref{fig:pi_over_8}.
Thus in this protocol, the technological requirements for preparing $\Ket{0}$ and $\Ket{H}$ are at the same level.

\end{document}